\documentclass[preprint,aps,12pt,showpacs,nofootinbib,tightenlines]{revtex4}
\usepackage{amsmath}
\usepackage{amssymb}
\usepackage{epsfig}
\usepackage{graphicx}
\begin{document}

\newcommand{\beq}{\begin{eqnarray}}
\newcommand{\eeq}{\end{eqnarray}}
\newcommand{\non}{\nonumber\\ }

\newcommand{\acp}{{\cal A}_{CP}}
\newcommand{\etap}{\eta^{(\prime)} }
\newcommand{\etapr}{\eta^\prime }
\newcommand{\jpsi}{ J/\Psi }
\newcommand{\kst} {K^{*0}}
\newcommand{\kstb}{\overline{K}^{*0}}

\newcommand{\pb}{\phi_{B}}
\newcommand{\pjL}{\phi_{2}^L}
\newcommand{\pjt}{\phi_{2}^t}
\newcommand{\pjT}{\phi_{2}^T}
\newcommand{\pjV}{\phi_{2}^V}
\newcommand{\p}{\phi_{3}}
\newcommand{\pms}{\phi_{3}^s}
\newcommand{\pmt}{\phi_{3}^t}
\newcommand{\pmv}{\phi_{3}^v}
\newcommand{\pmT}{\phi_{3}^T}
\newcommand{\pma}{\phi_{3}^a}
\newcommand{\fb}{f_{B_s} }
\newcommand{\fpi}{f_{\pi} }
\newcommand{\fj}{f_{J/\Psi} }
\newcommand{\fetap}{f_{\eta'} }
\newcommand{\rpi}{r_{\pi} }
\newcommand{\rp}{r_{3} }
\newcommand{\rj}{r_{2} }
\newcommand{\mb}{m_{B_s} }
\newcommand{\mop}{m_{0\pi} }
\newcommand{\moe}{m_{0\eta} }
\newcommand{\moep}{m_{0\eta'} }

\newcommand{\psl}{ P \hspace{-2.8truemm}/ }
\newcommand{\nsl}{ n \hspace{-2.2truemm}/ }
\newcommand{\vsl}{ v \hspace{-2.2truemm}/ }
\newcommand{\epsl}{\epsilon \hspace{-1.8truemm}/\,  }

\def \epjc{ Eur. Phys. J. C }
\def \jpg{  J. Phys. G }
\def \npb{  Nucl. Phys. B }
\def \plb{  Phys. Lett. B }
\def \pr{  Phys. Rep. }
\def \prd{  Phys. Rev. D }
\def \prl{  Phys. Rev. Lett.  }
\def \zpc{  Z. Phys. C  }
\def \jhep{ J. High Energy Phys.  }
\def \ijmpa { Int. J. Mod. Phys. A }

\title{ The pure annihilation type $B_c \to M_2 M_3$ decays in the perturbative QCD approach}
\author{Xin Liu$^a$\footnote{liuxin.physics@gmail.com},
Zhen-Jun Xiao$^a$\footnote{xiaozhenjun@njnu.edu.cn} and Cai-Dian
L\"u$^{(a,b)}$\footnote{lucd@ihep.ac.cn} } 
\affiliation{ a. Department of Physics and
Institute of Theoretical Physics, Nanjing Normal University,
Nanjing, Jiangsu 210046, P.R. China\footnote{Mailing address}\\
b. Institute of High Energy Physics and Theoretical Physics Center
for Science Facilities, CAS, P.O.Box 918(4), Beijing 100049,   P.R.
China } 
\date{\today}
\begin{abstract}
In the standard model the two-body charmless hadronic $B_c$ meson decays
can occur via annihilation diagrams only.
In this work, we studied the $B_c \to  PP, PV/VP,VV$ decays
by employing the perturbative QCD (pQCD) factorization approach.
From our calculations, we find that
(a) the pQCD predictions for the branching ratios of the considered $B_c$ decays
are in the range of $10^{-6}$ to $10^{-8}$;
(b) for $B_c \to PV/VP, VV$ decays, the branching ratios of  $\Delta S= 0$ decays
are much larger than those of $\Delta S =1$ ones because the different
Cabibbo-Kobayashi-Maskawa(CKM) factors are involved;
(c) analogous to $B \to K \eta^{(\prime)}$ decays, we find
$Br(B_c \to K^+ \eta^\prime) \sim 10 \times Br(B_c \to K^+ \eta)$, which
can be understood   by the destructive and constructive interference
between the $\eta_q$ and $\eta_s$ contribution to
the $B_c \to K^+ \eta$ and $B_c \to K^+ \eta^\prime$ decay;
(d) the longitudinal polarization fractions of $B_c \to VV$ decays are in the range of
$86\%-95\%$ and play the dominant role;
and (e) there is no CP-violating asymmetries for the considered $B_c$ decays
because only one type tree operators involved.
\end{abstract}


\maketitle

\section{Introduction}

In 1998, a new stage of $B_c$ physics began because of the first observation of
the meson $B_c$ at Tevatron ~\cite{cdf}. For $B_c$ meson, one can study
the two heavy flavors b and c in a meson simultaneously.
From an experimental point of view, more detailed information about $B_c$ meson
can be obtained  at the Large Hadron Collider (LHC) experiment.
The LHC is scheduled to start to run in this
month, where the $B_c$ meson could be produced abundantly.
The $B_c$ meson decays may provide windows for testing the
predictions of the standard model(SM) and can shed light on new physics(NP)
scenarios  beyond the SM.

From a theoretical point of view~\cite{nb04:bcre},
the non-leptonic decays of $B_c$ meson  are the most complicated decays due to
its heavy-heavy nature and the participation of strong interaction, which
complicate the extraction of  parameters in SM, but they also provide
great opportunities to study the perturbative
and nonperturbative QCD, final state interactions, etc.
The non-leptonic $B_c$ weak decays have been widely studied for example in
Refs.~\cite{nb04:bcre,Bc10,Gouz,KKL,IKS06,EFG03D,CC94,LC97,AMV,CD00,
HNV,DW,DSV,DSV2,Giri,KisJPG,LL08,CDL,FKS04,FW00,KPS,CMM,VAS,DD99,
Mas92,WK92,SYDM,IKP,HS84,yu09:epjc,cj09} by employing the Naive factorization
approach(NFA)~\cite{nfa}, the QCD factorization approach(QCDF)~\cite{qcdf},
the perturbative QCD (pQCD) approach~\cite{Keum01:kpi,Lu01:pipi,Li03:ppnp}
and other approaches and/or methods.

In this paper we focus on the two-body non-leptonic charmless decays
$B_c \to PP, PV/VP, VV$ (here P and V stands for the light
pseudo-scalar and vector mesons), which can occur through the weak
annihilation diagrams only.
The size of annihilation contributions is an important issue in $B$ physics.
Indeed, the two-body charmless $B_c$ decays considered here are
rather different from  those $B_c\to J/\psi P(V)$ decays where
the initial $c$ quark behaves as a spectator.

Recently, the two-body non-leptonic charmless  $B_c \to M_2 M_3$
\footnote{For the sake of simplicity, we will use $M_2$ and $M_3$ to denote the two
final state light mesons respectively, unless otherwise stated. } decays
have been studied by using the $SU(3)$ flavor symmetry or
by employing the QCD factorization approach \cite{ekou09:ncbc}.
The authors in Ref.~\cite{ekou09:ncbc} provided two different estimates
for non-leptonic charmless $B_c$ decays. But their predictions for the
branching ratios of $B_c \to \phi K^+, \overline{K}^{*0} K^+$ decays
in the QCDF are much smaller (a factor of 10) than those obtained by using the
SU(3) flavor symmetry.
So large discrepancies among the theoretical predictions for the branching ratios
indicate clearly that it is very necessary to make more studies for these kinds
of $B_c$ decays by employing other different approaches, in order to
understand these decays better and provide the theoretical support for the
related experimental studies.

In this paper, we will calculate the branching ratios and the polarization
fractions of thirty $B_c \to PP, PV/VP, VV$ decays
by employing the low energy effective Hamiltonian~\cite{Buras96:weak}
and the pQCD factorization approach. By keeping the transverse momentum
$k_T$ of the quarks, the pQCD approach
is free of endpoint singularity
and the Sudakov formalism makes it more self-consistent. It is worth
of mentioning  that one can do the quantitative calculations of the
annihilation type diagrams in the pQCD approach.
The importance of annihilation contributions has already
been tested in the previous predictions of branching ratios of pure
annihilation $B \to D_s K$ decays~\cite{Lu03:dsk}, direct CP
asymmetries of $B^0 \to \pi^+\pi^-$, $K^+\pi^-$
decays~\cite{Keum01:kpi,Lu01:pipi,Hong06:direct} and in the
explanation of $B\to \phi K^*$ polarization
problem~\cite{Li05:kphi,Gritsan07:kphi}, which indicate that the
pQCD approach is a reliable method to deal with annihilation diagrams.

The paper is organized as follows. In Sec.~\ref{sec:1}, we present
the formalism and wave functions of the considered $B_c$ meson decays.
Then we perform the perturbative calculations for considered
decay channels with pQCD approach in Sec.~\ref{sec:2}.
The numerical results and phenomenological analysis are given in
Sec.~\ref{sec:3}. Finally, Sec.~\ref{sec:sum} contains the main
conclusions and a short summary.

\section{Formalism and wave functions}\label{sec:1}

\subsection{Formalism}

Since the b quark is rather heavy, we work in the frame with the
$B_c$ meson at rest, i.e., with the $B_c$ meson momentum
$P_1=\frac{m_{B_c}}{\sqrt{2}}(1,1,{\bf 0}_T)$ in the light-cone
coordinates. For the non-leptonic charmless $B_c \to M_2 M_3$
decays, assuming that the $M_2$ ($M_3$) meson moves in the plus
(minus) $z$ direction carrying the momentum $P_2$ ($P_3$) and the
polarization vector $\epsilon_2$ ($\epsilon_3$)(if $M_{2(3)}$ are
the vector mesons). Then the two final state meson momenta can be
written as
\beq
     P_2 =\frac{m_{B_c}}{\sqrt{2}} (1-r_3^2,r_2^2,{\bf 0}_T), \quad
     P_3 =\frac{m_{B_c}}{\sqrt{2}} (r_3^2,1-r_2^2,{\bf 0}_T),
\eeq
respectively, where $r_2=m_{M_2}/m_{B}$,and
$r_3=m_{M_3}/m_{B}$. When $M_2,M_3$ are the vector mesons, the
longitudinal polarization vectors, $\epsilon_2^L$ and
$\epsilon_3^L$, can be given by
\beq
\epsilon_2^L =\frac{m_{B_c}}{\sqrt{2}m_{M_2}} (1-r_3^2, -r_2^2,{\bf 0}_T), \quad
\epsilon_3^L = \frac{m_{B_c}}{\sqrt{2}m_{M_3}} (-r_3^2, 1-r_2^2,{\bf0}_T).
\eeq
The transverse ones are parameterized as $\epsilon_2^T =
(0, 0,1_T)$, and $\epsilon_3^T = (0, 0,1_T)$. Putting the (light-)
 quark momenta in $B_c$, $M_2$ and $M_3$ mesons as $k_1$,
$k_2$, and $k_3$, respectively, we can choose
\beq
k_1 = (x_1P_1^+,0,{\bf k}_{1T}), \quad k_2 = (x_2 P_2^+,0,{\bf k}_{2T}), \quad
k_3 = (0, x_3 P_3^-,{\bf k}_{3T}).
\eeq
Then, for $B_c \to M_2 M_3$
decays, the
integration over $k_1^-$, $k_2^-$, and $k_3^+$ 
will conceptually lead to the decay amplitudes in the pQCD approach,
\beq
{\cal A}(B_c \to M_2 M_3) &\sim &\int\!\! d x_1 d x_2 d x_3 b_1
d b_1 b_2 d b_2 b_3 d b_3 \non && \cdot \mathrm{Tr} \left [ C(t)
\Phi_{B_c}(x_1,b_1) \Phi_{M_2}(x_2,b_2) \Phi_{M_3}(x_3, b_3) H(x_i,
b_i, t) S_t(x_i)\, e^{-S(t)} \right ]\;
\label{eq:a2}
\eeq
where $b_i$ is the conjugate space coordinate of $k_{iT}$, and $t$ is the
largest energy scale in function $H(x_i,b_i,t)$. The large
logarithms $\ln (m_W/t)$ are included in the Wilson coefficients
$C(t)$. The large double logarithms ($\ln^2 x_i$) are summed by the
threshold resummation ~\cite{Li02:resum}, and they lead to
$S_t(x_i)$ which smears the end-point singularities on $x_i$. The
last term, $e^{-S(t)}$, is the Sudakov form factor which suppresses
the soft dynamics effectively ~\cite{Li98:soft}. Thus it makes the
perturbative calculation of the hard part $H$ applicable at
intermediate scale, i.e., $m_{B_c}$ scale. We will calculate
analytically the function $H(x_i,b_i,t)$ for the considered decays
at leading order(LO) in $\alpha_s$ expansion and give the convoluted
amplitudes in next section.

For these considered decays, the related weak effective
Hamiltonian $H_{eff}$~\cite{Buras96:weak} can be written as
\beq
\label{eq:heff} H_{eff} = \frac{G_{F}}
{\sqrt{2}} \, \left[ V_{cb}^* V_{uD} \left (C_1(\mu) O_1(\mu) +
C_2(\mu) O_2(\mu) \right) \right] \;\label{heff} ,
\eeq
with the single tree operators,
\beq
O_1 &=& \bar u_\beta \gamma^\mu (1-\gamma_5) D_\alpha \bar c_\beta \gamma^\mu
(1- \gamma_5) b_\alpha \; , \non
O_2 &=& \bar u_\beta \gamma^\mu (1- \gamma_5) D_\beta
\bar c_\alpha \gamma^\mu (1- \gamma_5) b_\alpha \; ,
\eeq
where $V_{cb}, V_{uD}$ are the CKM matrix
elements, "D" denotes the light down quark $d$ or $s$ and
$C_i(\mu)$ are Wilson coefficients at the renormalization scale
$\mu$. For the Wilson coefficients $C_{1,2}(\mu)$, we will also
use the leading order (LO) expressions, although the
next-to-leading order calculations already exist in the
literature~\cite{Buras96:weak}. This is the consistent way to
cancel the explicit $\mu$ dependence in the theoretical formulae.
For the renormalization group evolution of the Wilson coefficients
from higher scale to lower scale, we use the formulae as given in
Ref.~\cite{Lu01:pipi} directly.

\subsection{Wave Functions}\label{ssec:wf}

In order to calculate the decay amplitude,
we should choose the proper wave functions of the heavy $B_c$ and light mesons.
In principle there are two Lorentz structures in the $B_{u,d,s}$ or
$B_c$ meson wave function. One should consider both of them in calculations.
However, since the contribution induced by one Lorentz structure is numerically
small~\cite{luy03:form,ali07:bspv} and can be neglected approximately,
we only consider the contribution from the first Lorentz structure.
\beq
\Phi_{B_c} (x) &=& \frac{i}{\sqrt{2 N_c}}\left[ (\psl  + M_{B_c}) \gamma_5
\phi_{B_c}(x) \right]_{\alpha\beta}\;.
\eeq
Since $B_c$ meson consists of two heavy quarks and $m_{B_c} \simeq m_b+m_c$, the
distribution amplitude $\phi_{B_c}$ would be close to
$\delta(x-m_c/m_{B_c})$ in the non-relativistic limit. We therefore
adopt the non-relativistic approximation form of $\phi_{B_c}$ as~\cite{CDL,SYDM},
\beq
\phi_{B_c}(x) &=& \frac{f_{B_c}}{2 \sqrt{2 N_c}} \delta (x-
m_c/m_{B_c})\;,
\eeq
where $f_{B_c}$ and $N_c$ are the decay constant of $B_c$ meson and
the color number, respectively.

For the pseudoscalar meson(P), the wave function can generally be
defined as,
\beq
\Phi_P(x) &=& \frac{i}{\sqrt{2 N_c}} \gamma_5
\left\{\psl \phi_P^A(x)+ m_0^P \phi_P^P(x) + m_0^P (\nsl \vsl -1)
\phi_P^T(x)\right\}_{\alpha\beta}
\eeq
where $\phi_P^{A,P,T}$ and $m_0^P$ are the distribution amplitudes and chiral scale parameter
of the pseudoscalar mesons respectively, while $x$ denotes the momentum
fraction carried by quark in the meson, and $n=(1,0,{\bf 0}_T)$
and $v=(0,1,{\bf 0}_T)$ are dimensionless light-like unit vectors.

For the wave functions of vector mesons, one
longitudinal(L) and two transverse(T) polarizations are involved, and
 can be written as,
\beq
\Phi^L_V(x)&=& \frac{1}{\sqrt{2 N_c}} \left\{
M_V \epsl_V^{*L} \phi_V(x) + \epsl^{*L}_V \psl \phi_V^t(x)+ M_V
\phi_V^s(x)\right\}_{\alpha\beta} \;, \\
\Phi^T_V(x)&=&\frac{1}{\sqrt{2 N_c}} \left\{ M_V \epsl_V^{*T} \phi_V^v(x) +
\epsl^{*T}_V \psl \phi_V^T(x)+ M_V i \epsilon_{\mu\nu\rho\sigma}
\gamma_5 \gamma^\mu \epsilon_T^{*\nu} n^\rho v^\sigma
\phi_V^a(x)\right\}_{\alpha\beta} \;,
\eeq
where $\epsilon_V^{L(T)}$ denotes the longitudinal(transverse) polarization
vector of vector mesons, satisfying $P \cdot \epsilon=0$ in each polarization.
We here adopt the convention $\epsilon^{0123}=1$ for the Levi-Civita
tensor $\epsilon^{\mu\nu\alpha\beta}$. For the distribution
amplitudes of pseudoscalar $\phi_P^{A,P,T}$, and longitudinal and
transverse polarization, $\phi_V^{,t,s}$ and $\phi_V^{v,T,a}$,
which will be presented in Appendix~\ref{sec:app1}.

\section{perturbative calculations in pQCD} \label{sec:2}

\begin{figure}[t,b]
\vspace{-0.5cm} \centerline{\epsfxsize=13 cm \epsffile{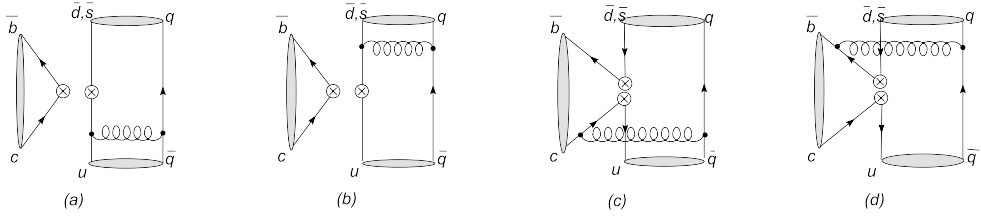}}
\vspace{0.2cm} \caption{Typical Feynman diagrams for two-body
non-leptonic charmless $B_c$ decays.}
 \label{fig:fig1}
\end{figure}

From the effective Hamiltonian~(\ref{heff}), there are 4 types of
diagrams contributing to the $B_c \to M_2 M_3$ decays as
illustrated in Fig.~\ref{fig:fig1}, which result in the Feynman
decay amplitudes $F_{fa}^{M_2 M_3}$ and $M_{na}^{M_2 M_3}$,  where
the subscripts ${fa}$ and ${na}$  are the abbreviations of
factorizable and non-factorizable annihilation contributions, respectively.
Operators $O_{1,2}$ are $(V-A)(V-A)$ currents, we
therefore can combine all contributions from these diagrams and
obtain the total decay amplitude as,
\beq
{\cal A}(B_c \to  M_2
M_3) &=& V_{cb}^* V_{uD} \left\{f_{B_c} F_{fa}^{M_2 M_3}a_1 +
M_{na}^{M_2 M_3} C_1 \right\} \; ,
\label{eq:amt}
\eeq
where $a_1=C_1/3+C_2$.  In the next three subsections we will give the
explicit expressions of $F_{fa}^{M_2M_3}$, $M_{na}^{ M_2 M_3}$ and
the decay amplitude ${\cal A}(B_c \to  M_2 M_3)$ for $B_c\to M_2 M_3$
decays: including eight $B_c \to P P$, fifteen $B_c \to P V$ or
$B_c \to V P$, and seven $B_c \to V V$ decay modes.

\subsection{  $B_c \to PP$ decays}
In this section, we will present the factorization formulae for
eight non-leptonic charmless $B_c \to P P$ decays. From the first two
diagrams of Fig.~\ref{fig:fig1}, i.e., (a) and (b), by perturbative
QCD calculations, we obtain the decay amplitude
for factorizable annihilation contributions as follows,
\beq
F_{fa}^{PP} &=&- 8 \pi C_F m_{B_c}^2 \int_0^1 d x_{2} dx_{3}\,
\int_{0}^{\infty} b_2 db_2 b_3 db_3\, \non && \times
\left\{h_{fa}(1-x_{3},x_{2},b_{3},b_{2})E_{fa}(t_{a}) \left[x_{2}
\phi_{2}^A(x_2)\phi^A_{3}(x_3)+2 r_0^2 r_0^3\phi_{3}^P(x_3)
\right.\right. \non && \left.\left. \times
 \left((x_2 + 1)\phi^{P}_{2}(x_2)+ (x_2 -
1)\phi^{T}_{2}(x_{2})\right)\right]+h_{fa}(x_{2},1-x_{3},b_{2},b_{3})E_{fa}(t_{b})\right.
\non && \left.
 \times\left[ (x_3
-1) \phi_{2}^A(x_2) \phi_{3}^A(x_3) + 2 r_0^2 r_0^3
\phi_{2}^P(x_2) \left( (x_3 -2)\phi_{3}^P(x_3)- x_3
\phi_{3}^T(x_3)\right)\right] \right\}\;, \label{eq:ab}
\eeq
where $\phi_{2(3)}$ corresponding to the distribution amplitudes of
mesons $M_{2(3)}$, $r_0^{2(3)}= m_{0}^{M_2(M_3)}/m_{B_c}$, and
$C_F=4/3$ is a color factor. In Eq.~(\ref{eq:ab}), the terms
proportional to $({r_0^{2(3)}})^2$ have been neglected because
they are small indeed, $\max({r_0^{2(3)}})^2 \leq 7\% $. The
function $h_{fa}$, the scales $t_i$ and $E_{fa}(t)$ can be found
in Appendix~\ref{sec:app2}.

For the non-factorizable diagrams (c) and (d), all three meson wave
functions are involved. The integration of $b_3$ can be performed
using $\delta$ function $\delta(b_3-b_2)$, leaving only integration
of $b_1$ and $b_2$. The corresponding decay amplitude is
\beq
 M_{na}^{PP} &=& -\frac{16 \sqrt{6}}{3}\pi C_F m_{B_c}^2
\int_{0}^{1}d x_{2}\,d x_{3}\,\int_{0}^{\infty} b_1d b_1 b_2d b_2\,
 \non && \times \left\{h_{na}^{c}(x_2,x_3,b_1,b_2) E_{na}(t_c)
\left[(r_c - x_3 +1) \phi_{2}^A(x_2)\phi_{3}^A (x_{3})+ r_0^2
r_0^3\left(\phi_{2}^P(x_2) \right.\right. \right.\non && \left. \left.
\left. \times ((3 r_c + x_2 -x_3 +1) \phi_{3}^P(x_3)-(r_c -x_2 -x_3
+1)\phi_{3}^T(x_3))+\phi_{2}^T(x_2)\right.\right.\right.
\non &&
\left.\left. \left. \times ((r_c-x_2 -x_3 +1) \phi_{3}^P(x_3)+(r_c
-x_2 +x_3 -1)\phi_{3}^T(x_3))\right)\right]-E_{na}(t_d) \right.
\non
&& \left.
\times\left[ (r_b + r_c +x_2 -1) \phi_{2}^A(x_2) \phi
_{3}^A(x_3) + r_0^2 r_0^3 \left(\phi_{2}^P(x_2)((4 r_b
 +r_c +x_2 -x_3\right.\right.\right.\non
&& \left.
\left.\left. -1)\phi_{3}^P(x_3)-(r_c + x_2 +x_3 -1)\phi_{3}^T(x_3))
+\phi_{2}^T(x_2)((r_c + x_2 +x_3 -1)\right.\right.\right. \non
&&
\left.\left.\left. \times\phi_{3}^P(x_3)-(r_c +x_2 -x_3 -1)
\phi_{3}^T(x_3))\right)\right]
h_{na}^{d}(x_2,x_3,b_1,b_2)\right\}\;,\label{eq:cd}
\eeq
where $r_{b(c)}= m_{b(c)}/m_{B_c}$.

For the ¡°$\eta-\eta^\prime$" system, there exist two popular
mixing basis: the octet-singlet basis and the quark-flavor basis
\cite{fk98:etap,ef05:etap}. Here we  use  the quark-flavor basis
\cite{fk98:etap} and define
\beq
\eta_q=(u\bar{u} +
d\bar{d})/\sqrt{2}, \qquad \eta_s=s\bar{s}. \label{eq:qfbasis}
\eeq
The physical states $\eta$ and $\eta^\prime$ are related to
$\eta_q$ and $\eta_s$ through a single mixing angle $\phi$,
\beq
\left( \begin{array}{c} \eta\\ \eta^\prime \\ \end{array} \right ) &=&
U(\phi) \left( \begin{array}{c}
 \eta_q\\ \eta_s \\ \end{array} \right ) =
  \left( \begin{array}{cc}
 \cos{\phi} & -\sin{\phi} \\
 \sin{\phi} & \cos\phi \end{array} \right )
 \left( \begin{array}{c}  \eta_q\\ \eta_s \\ \end{array} \right ).
 \label{eq:e-ep}
\eeq
We assume that the distribution amplitudes of $\eta_q$ and
$\eta_s$ are  the same
as the distribution amplitudes of $\pi$, except for the different
decay constants and the chiral scale parameters. The three input
parameters $f_q$, $f_s$ and $\phi$ in the quark-flavor basis have
been extracted from various related experiments
\cite{fk98:etap,ef05:etap}
\beq
f_q = (1.07\pm 0.02) f_\pi, \quad
f_s = (1.34 \pm 0.06) f_\pi, \quad \phi = 39.3^\circ \pm 1.0^\circ.
\label{eq:e}
\eeq
The chiral enhancement factors are chosen as
\beq
  m_0^{\eta_q}&\equiv&
  \frac{m_{qq}^2}{2m_q}=\frac{1}{2m_q}[m^2_\eta\cos^2\phi+
  m_{\eta'}^2\sin^2\phi-\frac{\sqrt 2f_s}{f_q}(m_{\eta'}^2-m_\eta^2)\cos\phi\sin\phi],\\
  m_0^{\eta_s}&\equiv&
  \frac{m_{ss}^2}{2m_s}=\frac{1}{2m_s}[m^2_{\eta'}\cos^2\phi+
  m_{\eta}^2\sin^2\phi-\frac{f_q}{\sqrt
  2f_s}(m_{\eta'}^2-m_\eta^2)\cos\phi\sin\phi].
\eeq
In the numerical calculations, we will use these mixing parameters as
inputs. It is worth of mentioning that the effects of possible gluonic
component of $\eta^\prime$ meson will not considered here since it
is small in size \cite{xiao06,0609165,0703187}.

Based on Eqs.(\ref{eq:amt}) and (\ref{eq:ab},\ref{eq:cd}),
we can write down the total decay amplitudes for eight $B_c \to PP$
decays easily,
\beq
{\cal A}(B_c \to \pi^+ \pi^0) &=& V_{cb}^* V_{ud} \left\{[f_{B_c} F_{fa}^{\pi^+
\pi_{\bar u u}^0}a_1 + M_{na}^{\pi^+ \pi_{\bar u u}^0} C_1 ] \right. \non && \left. -[f_{B_c}
F_{fa}^{\pi_{\bar d d}^0\pi^+}
a_1 + M_{na}^{\pi_{\bar d d}^0\pi^+ } C_1 ]  \right\}=0\;,\label{eq:pipi0} \\
{\cal A}(B_c \to \pi^+ \eta) &=& V_{cb}^* V_{ud} \left\{[f_{B_c}
F_{fa}^{\pi^+ \eta_{\bar u u}}a_1 + M_{na}^{\pi^+ \eta_{\bar u u}} C_1 ] \right. \non && \left.
+ [f_{B_c}
F_{fa}^{\eta_{\bar d d}\pi^+ }a_1 + M_{na}^{\eta_{\bar d d}\pi^+ } C_1 ]\right\}\cos\phi\;,\\
{\cal A}(B_c \to \pi^+ \eta') &=& V_{cb}^*
V_{ud} \left\{[f_{B_c} F_{fa}^{\pi^+
\eta_{\bar u u}}a_1 + M_{na}^{\pi^+ \eta_{\bar u u}} C_1] \right. \non && \left.
+ [f_{B_c} F_{fa}^{\eta_{\bar d d} \pi^+}a_1 + M_{na}^{\eta_{\bar d d} \pi^+ } C_1] \right\}\sin\phi\;,\\
{\cal A}(B_c \to \overline{K}^0 K^+) &=& V_{cb}^* V_{ud}
\left\{f_{B_c} F_{fa}^{\overline{K}^0 K^+}a_1 +
M_{na}^{\overline{K}^0 K^+} C_1 \right\}\;,\eeq
\beq
{\cal A}(B_c \to  K^+ \pi^0) &=& V_{cb}^* V_{us} \left\{f_{B_c}
F_{fa}^{K^+\pi^0}a_1 + M_{na}^{K^+\pi^0} C_1 \right\}\;,\\
{\cal A}(B_c \to  K^0 \pi^+) &=&  \sqrt{2} {\cal A}(B_c \to  K^+ \pi^0)\;,\\
{\cal A}(B_c \to K^+ \eta) &=& V_{cb}^* V_{us} \left\{f_{B_c}
\left [F_{fa}^{K^+ \eta_q}\cos\phi-F_{fa}^{\eta_s K^+ }\sin\phi \right ] a_1\right.
\non && \left.  +
\left [M_{na}^{K^+ \eta_q}\cos\phi-M_{na}^{\eta_s K^+ }\sin\phi \right ] C_1 \right\}\;,\\
{\cal A}(B_c \to K^+ \eta') &=& V_{cb}^* V_{us} \left\{f_{B_c}
\left [F_{fa}^{K^+ \eta_q}\sin\phi+F_{fa}^{\eta_s K^+ }\cos\phi \right ]a_1\right.
\non &&
\left.  + \left [M_{na}^{K^+ \eta_q}\sin\phi+M_{na}^{\eta_s K^+
}\cos\phi \right ] C_1 \right\}\;. \label{eq:ketap}
\eeq

\subsection{  $B_c \to PV , VP$ decays}
By following the same procedure as stated in the above subsection,
we can obtain the analytic decay amplitudes for $B_c \to PV, VP$ decays,
\beq
F_{fa}^{PV} &=&8 \pi C_F m_{B_c}^2 \int_0^1 d x_{2} dx_{3}\,
\int_{0}^{\infty} b_2 db_2 b_3 db_3\, \non && \times
\left\{h_{fa}(1-x_{3},x_{2},b_{3},b_{2})E_{fa}(t_{a}) \left[x_{2}
\phi_{2}^A(x_2)\phi_{3}(x_3)-2 r_0^2 r_3\phi_{3}^s(x_3) \right.\right.
\non && \left.\left. \times
 \left((x_2 + 1)\phi^{P}_{2}(x_2)+ (x_2 -
1)\phi^{T}_{2}(x_{2})\right)\right]+
h_{fa}(x_{2},1-x_{3},b_{2},b_{3})E_{fa}(t_{b})\right.
\non && \left.
 \times\left[ (x_3
-1) \phi_{2}^A(x_2) \phi_{3}(x_3) - 2 r_0^2 r_3 \phi_{2}^P(x_2) \left(
(x_3 -2)\phi_{3}^s(x_3)- x_3 \phi_{3}^t(x_3)\right)\right]
\right\}\;, \label{eq:ab1}
\\
M_{na}^{PV} &=& \frac{16 \sqrt{6}}{3}\pi C_F
m_{B_c}^2 \int_{0}^{1}d x_{2}\,d x_{3}\,\int_{0}^{\infty} b_1d b_1
b_2d b_2\,
 \non && \times \left\{h_{na}^{c}(x_2,x_3,b_1,b_2) E_{na}(t_c)
\left[(r_c - x_3 +1) \phi_{2}^A(x_2)\phi_{3}(x_{3})- r_0^2
r_3\left(\phi_{2}^P(x_2) \right.\right. \right.\non && \left. \left.
\left. \times ((3 r_c + x_2 -x_3 +1) \phi_{3}^s(x_3)-(r_c -x_2 -x_3
+1)\phi_{3}^t(x_3))+\phi_{2}^T(x_2)\right.\right.\right. \non &&
\left.\left. \left. \times ((r_c-x_2 -x_3 +1) \phi_{3}^s(x_3)+(r_c
-x_2 +x_3 -1)\phi_{3}^t(x_3))\right)\right]-E_{na}(t_d) \right. \non
&& \left. \times\left[ (r_b + r_c +x_2 -1) \phi_{2}^A(x_2) \phi
_{3}(x_3) - r_0^2 r_3 \left(\phi_{2}^P(x_2)((4 r_b
 +r_c +x_2 -x_3\right.\right.\right.\non && \left.
\left.\left. -1)\phi_{3}^s(x_3)-(r_c + x_2 +x_3 -1)\phi_{3}^t(x_3))
+\phi_{2}^T(x_2)((r_c + x_2 +x_3 -1)\right.\right.\right. \non &&
\left.\left.\left. \times\phi_{3}^s(x_3)-(r_c +x_2 -x_3 -1)
\phi_{3}^t(x_3))\right)\right]
h_{na}^{d}(x_2,x_3,b_1,b_2)\right\}\;,\label{eq:cd1}\eeq
\beq
F_{fa}^{VP}
&=&8 \pi C_F m_{B_c}^2 \int_0^1 d x_{2} dx_{3}\, \int_{0}^{\infty}
b_2 db_2 b_3 db_3\, \non && \times
\left\{h_{fa}(1-x_{3},x_{2},b_{3},b_{2})E_{fa}(t_{a}) \left[x_{2}
\phi_{2}(x_2)\phi^A_{3}(x_3)+2 r_2 r_0^3\phi_{3}^P(x_3) \right.\right.
\non && \left.\left. \times
 \left((x_2 + 1)\phi^{s}_{2}(x_2)+ (x_2 -
1)\phi^{t}_{2}(x_{2})\right)\right]+h_{fa}(x_{2},1-x_{3},b_{2},b_{3})E_{fa}(t_{b})\right.
\non && \left.
 \times\left[ (x_3
-1) \phi_{2}(x_2) \phi_{3}^A(x_3) + 2 r_2 r_0^3 \phi_{2}^s(x_2) \left(
(x_3 -2)\phi_{3}^P(x_3)- x_3 \phi_{3}^T(x_3)\right)\right]
\right\}\;, \label{eq:ab2}\\
M_{na}^{VP} &=& \frac{16 \sqrt{6}}{3}\pi C_F
m_{B_c}^2 \int_{0}^{1}d x_{2}\,d x_{3}\,\int_{0}^{\infty} b_1d b_1
b_2d b_2\,
 \non && \times \left\{h_{na}^{c}(x_2,x_3,b_1,b_2) E_{na}(t_c)
\left[(r_c - x_3 +1) \phi_{2}(x_2)\phi_{3}^A (x_{3})+ r_2
r_0^3\left(\phi_{2}^s(x_2) \right.\right. \right.\non && \left. \left.
\left. \times ((3 r_c + x_2 -x_3 +1) \phi_{3}^P(x_3)-(r_c -x_2 -x_3
+1)\phi_{3}^T(x_3))+\phi_{2}^t(x_2)\right.\right.\right. \non &&
\left.\left. \left. \times ((r_c-x_2 -x_3 +1) \phi_{3}^P(x_3)+(r_c
-x_2 +x_3 -1)\phi_{3}^T(x_3))\right)\right]-E_{na}(t_d) \right. \non
&& \left. \times\left[ (r_b + r_c +x_2 -1) \phi_{2}(x_2) \phi
_{3}^A(x_3) + r_2 r_0^3 \left(\phi_{2}^s(x_2)((4 r_b
 +r_c +x_2 -x_3\right.\right.\right.\non && \left.
\left.\left. -1)\phi_{3}^P(x_3)-(r_c + x_2 +x_3 -1)\phi_{3}^T(x_3))
+\phi_{2}^t(x_2)((r_c + x_2 +x_3 -1)\right.\right.\right. \non &&
\left.\left.\left. \times\phi_{3}^P(x_3)-(r_c +x_2 -x_3 -1)
\phi_{3}^T(x_3))\right)\right]
h_{na}^{d}(x_2,x_3,b_1,b_2)\right\}\;,\label{eq:cd2}
\eeq

The total decay amplitudes of the fifteen $B_c \to PV, VP$ decays can therefore
be written as,
\beq
{\cal A}(B_c \to \pi^+ \rho^0) &=& V_{cb}^* V_{ud}
\left\{[f_{B_c} F_{fa}^{\pi^+\rho_{\bar u u}^0}a_1 + M_{na}^{\pi^+ \rho_{\bar u u}^0} C_1]
\right. \non && \left.
- [f_{B_c} F_{fa}^{\rho_{\bar d d}^0\pi^+}a_1 + M_{na}^{\rho_{\bar d d}^0\pi^+ } C_1] \right\}\;,\label{eq:pirho} \\
{\cal A}(B_c \to \pi^+ \omega) &=& V_{cb}^* V_{ud} \left\{
[f_{B_c}F_{fa}^{\pi^+ \omega_{\bar u u}}a_1 + M_{na}^{\pi^+ \omega_{\bar u u}} C_1] \right. \non && \left.
+ [f_{B_c}F_{fa}^{\omega_{\bar d d}\pi^+}a_1 + M_{na}^{\omega_{\bar d d}\pi^+ } C_1]
\right\}\;,
\\
{\cal A}(B_c \to \overline{K}^0 K^{*+}) &=& V_{cb}^* V_{ud}
\left\{f_{B_c} F_{fa}^{\overline{K}^0 K^{*+}}a_1 +
M_{na}^{\overline{K}^0 K^{*+}} C_1 \right\}\;,\eeq
\beq
{\cal A}(B_c \to  K^+ \rho^0) &=& V_{cb}^* V_{us} \left\{f_{B_c}
F_{fa}^{K^+ \rho^0}a_1 + M_{na}^{K^+ \rho^0} C_1 \right\}\;,\\
{\cal A}(B_c \to
 K^0 \rho^+) &=&  \sqrt{2} {\cal A}(B_c \to
 K^+ \rho^0)\;,\\
{\cal A}(B_c \to K^+ \omega) &=& V_{cb}^* V_{us} \left\{f_{B_c}
F_{fa}^{K^+ \omega}a_1 + M_{na}^{K^+ \omega} C_1 \right\}\;,
\eeq
\beq
{\cal A}(B_c \to \rho^+ \pi^0) &=& V_{cb}^* V_{ud} \left\{
[f_{B_c}F_{fa}^{\rho^+ \pi_{\bar u u}^0}a_1 + M_{na}^{\rho^+ \pi_{\bar u u}^0} C_1] \right. \non && \left.
- [f_{B_c}F_{fa}^{ \pi_{\bar d d}^0\rho^+}a_1 + M_{na}^{ \pi_{\bar d d}^0\rho^+} C_1]\right\}\;, \\
{\cal A}(B_c \to \rho^+ \eta) &=& V_{cb}^* V_{ud} \left\{
[f_{B_c}F_{fa}^{\rho^+ \eta_{\bar u u}}a_1 + M_{na}^{\rho^+\eta_{\bar u u}} C_1] \right. \non && \left.
+ [f_{B_c}F_{fa}^{ \eta_{\bar d d}\rho^+}a_1 + M_{na}^{\eta_{\bar d d}\rho^+} C_1] \right\}\cos\phi\;,\\
{\cal A}(B_c \to \rho^+ \eta') &=& V_{cb}^*
V_{ud} \left\{
[f_{B_c}F_{fa}^{\rho^+ \eta_{\bar u u}}a_1 + M_{na}^{\rho^+\eta_{\bar u u}} C_1] \right. \non && \left.
+ [f_{B_c}F_{fa}^{ \eta_{\bar d d}\rho^+}a_1 + M_{na}^{\eta_{\bar d d}\rho^+} C_1] \right\}\sin\phi\;,
\\
{\cal A}(B_c \to \overline{K}^{*0} K^+) &=& V_{cb}^* V_{ud}
\left\{f_{B_c} F_{fa}^{\overline{K}^{*0} K^+}a_1 +
M_{na}^{\overline{K}^{*0} K^+} C_1 \right\}\;,\eeq
\beq
{\cal A}(B_c \to  K^{*+} \pi^0) &=& V_{cb}^* V_{us} \left\{f_{B_c}
F_{fa}^{K^{*+}\pi^0}a_1 + M_{na}^{K^{*+}\pi^0} C_1 \right\}\;,\\
{\cal A}(B_c \to  K^{*0} \pi^+) &=&  \sqrt{2} {\cal A}(B_c \to
 K^{*+} \pi^0)\;,
 \eeq
\beq
{\cal A}(B_c \to K^{*+}\eta) &=& V_{cb}^* V_{us} \left\{f_{B_c}
\left [F_{fa}^{K^{*+}\eta_q}\cos\phi-F_{fa}^{\eta_s K^{*+}
}\sin\phi \right ]a_1\right. \non && \left.  +
\left [M_{na}^{K^{*+} \eta_q}\cos\phi-M_{na}^{\eta_s K^{*+} }\sin\phi \right ] C_1 \right\}\;,\\
{\cal A}(B_c \to K^{*+} \eta') &=& V_{cb}^* V_{us} \left\{f_{B_c}
\left [F_{fa}^{K^{*+} \eta_q}\sin\phi+F_{fa}^{\eta_s K^{*+}
}\cos\phi \right ]a_1\right. \non && \left.
+ \left [M_{na}^{K^{*+}\eta_q}\sin\phi+M_{na}^{\eta_s K^{*+} }\cos\phi \right ] C_1 \right\}\;,\\
{\cal A}(B_c \to
 \phi K^+) &=& V_{cb}^* V_{us} \left\{f_{B_c}
F_{fa}^{\phi K^+}a_1 + M_{na}^{\phi K^+} C_1 \right\}\;.
\label{eq:phik}
\eeq

\subsection{  $B_c \to VV$ decays}

There are three kinds of polarizations of a vector meson, namely,
longitudinal (L), normal (N), and transverse (T). The amplitudes for
a $B_c$ meson decay to two vector mesons are also characterized by
the polarization states of these vector mesons. The decay amplitudes
${\cal M}^{(\sigma)}$ in terms of helicities, for $B_c \to
V(P_2,\epsilon^*_2) V(P_3,\epsilon^*_3)$ decays, can be generally
described by
\beq
{\cal M}^{(\sigma)}
&=&\epsilon_{2\mu}^{*}(\sigma)\epsilon_{3\nu}^{*}(\sigma) \left[ a
\,\, g^{\mu\nu} + {b \over m_{M_2} m_{M_3}} P_1^\mu P_1^\nu + i{c
\over m_{M_2} m_{M_3}} \epsilon^{\mu\nu\alpha\beta} P_{2\alpha}
P_{3\beta}\right]\;,\non
&\equiv &m_{B_c}^{2}{\cal M}_{L}+m_{B_c}^{2}{\cal M}_{N}
\epsilon^{*}_{2}(\sigma=T)\cdot\epsilon^{*}_{3}(\sigma=T)
\non
&& +i{\cal
M}_{T}\epsilon^{\alpha \beta\gamma \rho}
\epsilon^{*}_{2\alpha}(\sigma)\epsilon^{*}_{3\beta}(\sigma)
P_{2\gamma }P_{3\rho }\; ,
\label{eq:msigma}
\eeq 
where the superscript $\sigma$ denotes the helicity
states of the two vector mesons with $L(T)$ standing for the longitudinal
(transverse) component. And the definitions of the amplitudes $
{\cal M}_{i}$ $(i=L,N,T)$ in terms of the Lorentz-invariant
amplitudes $a$, $b$ and $c$ are
\beq
m_{B_c}^2 \,\, {\cal M}_L &=& a
\,\, \epsilon_2^{*}(L) \cdot \epsilon_3^{*}(L) +{b \over m_{M_2}
m_{M_3}} \epsilon_{2}^{*}(L) \cdot P_3 \,\, \epsilon_{3}^{*}(L)
\cdot P_2\;, \non
m_{B_c}^2 \,\, {\cal M}_N &=& a \;,\label{id-rel}\non
m_{B_c}^2 \,\, {\cal M}_T &=& {c \over r_2\, r_3}\;. \label{eq:amp}
\eeq

We therefore will evaluate the helicity amplitudes ${\cal M}_L,
{\cal M}_N, {\cal M}_T$ based on the pQCD factorization approach,
respectively.

For every component of the polarization,
the corresponding Feynman amplitude can be written as the following form,
\beq
F^{L}_{fa} &=& 8 \pi C_F m_{B_c}^2 \int_0^1 d x_{2}
dx_{3}\, \int_{0}^{\infty} b_2 db_2 b_3 db_3\, \non && \times
\left\{ \left[x_{2} \phi_{2}(x_2)\phi_{3}(x_3) - 2 r_2 r_3
\left((x_2 + 1)\phi^{s}_{2}(x_2)+ (x_2 -
1)\phi^{t}_{2}(x_{2})\right)\right.\right.\non && \left.\left.
\times \phi_{3}^s(x_3)\right]E_{fa}(t_{a})
  h_{fa}(1-x_{3},x_{2},b_{3},b_{2})+E_{fa}(t_{b})
h_{fa}(x_{2},1-x_{3},b_{2},b_{3}) \right. \non && \left. \times
\left[ (x_3 -1) \phi_{2}(x_2) \phi_{3}(x_3) - 2 r_2 r_3
\phi_{2}^s(x_2) \left( (x_3 -2)\phi_{3}^s(x_3)- x_3
\phi_{3}^t(x_3)\right)\right] \right\}\;, \label{eq:abl}\\
M_{na}^{L} &=& \frac{16 \sqrt{6}}{3}\pi C_F m_{B_c}^2
\int_{0}^{1}d x_{2}\,d x_{3}\,\int_{0}^{\infty} b_1d b_1 b_2d b_2\,
 \non && \times \left\{E_{na}(t_c)\left[ (r_c - x_3 +1)
\phi_{2}(x_2)\phi_{3}(x_{3}) - r_2 r_3 \left(\phi_{2}^s(x_2)((3 r_c
+ x_2 -x_3 +1)  \right.\right.\right. \non && \left. \left. \left.
\times \phi_{3}^s(x_3)-(r_c -x_2 -x_3 +1)
\phi_{3}^t(x_3))+\phi_{2}^t(x_2)((r_c-x_2 -x_3 +1)
\phi_{3}^s(x_3)\right.\right.\right. \non && \left.\left. \left.
+(r_c -x_2 +x_3 -1)\phi_{3}^t(x_3))\right)\right]
h_{na}^{c}(x_2,x_3,b_1,b_2)-h_{na}^{d}(x_2,x_3,b_1,b_2)E_{na}(t_d)
\right. \non && \left.  \times \left[ (r_b + r_c +x_2 -1)
\phi_{2}(x_2) \phi _{3}(x_3) - r_2 r_3 \left(\phi_{2}^s(x_2)((4
r_b+r_c +x_2 -x_3 -1) \right.\right.\right.\non && \left.
\left.\left. \times \phi_{3}^s(x_3)-(r_c + x_2 +x_3
-1)\phi_{3}^T(x_3))+\phi_{2}^t(x_2)((r_c + x_2 +x_3
-1)\phi_{3}^s(x_3)\right.\right.\right. \non && \left. \left. \left.
-(r_c +x_2 -x_3 -1) \phi_{3}^t(x_3))\right)\right]
\right\}\;,\label{eq:cdl}
\eeq
\beq
F^{N}_{fa} &=& 8 \pi C_F m_{B_c}^2 \int_0^1 d x_{2} dx_{3}\,
\int_{0}^{\infty} b_2 db_2 b_3 db_3\,r_2 r_3 \non && \times \left\{
h_{fa}(1-x_{3},x_{2},b_{3},b_{2})E_{fa}(t_{a})\left[ (x_{2}+1 )
(\phi_{2}^a(x_2)\phi^a_{3}(x_3) +
\phi_{2}^v(x_2)\phi^v_{3}(x_3))\right.\right.\non && \left.\left. +
(x_2 - 1)
(\phi^{v}_{2}(x_{2})\phi_{3}^a(x_3)+\phi^{a}_{2}(x_{2})\phi_{3}^v(x_3))\right]
 +E_{fa}(t_{b})h_{fa}(x_{2},1-x_{3},b_{2},b_{3}) \right. \non && \left.  \times\left[(x_3
-2) (\phi_{2}^a(x_2) \phi_{3}^a(x_3)+\phi_{2}^v(x_2) \phi_{3}^v(x_3)
) - x_3
\left(\phi_{2}^a(x_2)\phi_{3}^v(x_3)+\phi_{2}^v(x_2)\phi_{3}^a(x_3)\right)\right]
 \right\}\;, \label{eq:abn} \\
M_{na}^{N} &=& \frac{32 \sqrt{6}}{3}\pi C_F m_{B_c}^2
\int_{0}^{1}d x_{2}\,d x_{3}\,\int_{0}^{\infty} b_1d b_1 b_2d b_2\,
r_2 r_3
 \non && \times \left\{r_c\left[
\phi_{2}^a(x_2)\phi_{3}^a(x_{3})+\phi_{2}^v(x_2)\phi_{3}^v (x_{3})
\right]E_{na}(t_c) h_{na}^{c}(x_2,x_3,b_1,b_2) \right. \non &&
\left.  -r_b\left[
\phi_{2}^a(x_2)\phi_{3}^a(x_{3})+\phi_{2}^v(x_2)\phi_{3}^v (x_{3})
\right]
E_{na}(t_d)h_{na}^{d}(x_2,x_3,b_1,b_2)\right\}\;,\label{eq:cdn}
\eeq
\beq
F^{T}_{fa} &=& 16 \pi C_F m_{B_c}^2 \int_0^1 d x_{2} dx_{3}\,
\int_{0}^{\infty} b_2 db_2 b_3 db_3\,r_2 r_3 \non && \times \left\{
h_{fa}(1-x_{3},x_{2},b_{3},b_{2})E_{fa}(t_{a})\left[ (x_{2}+1 )
(\phi_{2}^a(x_2)\phi^v_{3}(x_3) +
\phi_{2}^v(x_2)\phi^a_{3}(x_3))\right.\right.\non && \left.\left. +
(x_2 -
1)(\phi^{a}_{2}(x_{2})\phi_{3}^a(x_3)+\phi^{v}_{2}(x_{2})\phi_{3}^v(x_3))\right]
 +h_{fa}(x_{2},1-x_{3},b_{2},b_{3})E_{fa}(t_{b}) \right. \non && \left. \times  \left[(x_3
-2) (\phi_{2}^a(x_2) \phi_{3}^v(x_3)+\phi_{2}^v(x_2) \phi_{3}^a(x_3)
) - x_3
\left(\phi_{2}^a(x_2)\phi_{3}^a(x_3)+\phi_{2}^v(x_2)\phi_{3}^v(x_3)\right)\right]
 \right\}\;, \label{eq:abt}\\
M_{na}^{T} &=& \frac{64 \sqrt{6}}{3}\pi C_F m_{B_c}^2
\int_{0}^{1}d x_{2}\,d x_{3}\,\int_{0}^{\infty} b_1d b_1 b_2d b_2\,
r_2 r_3
 \non && \times \left\{r_c\left[
\phi_{2}^a(x_2)\phi_{3}^v(x_{3})+\phi_{2}^v(x_2)\phi_{3}^a (x_{3})
\right]E_{na}(t_c) h_{na}^{c}(x_2,x_3,b_1,b_2) \right. \non &&
\left.  -r_b\left[
\phi_{2}^a(x_2)\phi_{3}^v(x_{3})+\phi_{2}^v(x_2)\phi_{3}^a (x_{3})
\right]
E_{na}(t_d)h_{na}^{d}(x_2,x_3,b_1,b_2)\right\}\; .\label{eq:cdt}
\eeq

For seven $B_c \to V V$ decays, considering all the
polarization($H=L,N,T$) contributions and the Feynman
decay amplitudes as shown in Eqs.(\ref{eq:abl}-\ref{eq:cdt}), the total
decay amplitude of these channels can be obtained directly,
\beq
{\cal M}^{H}(B_c \to
\rho^+ \rho^0) &=& V_{cb}^* V_{ud} \left\{[f_{B_c} F_{fa;H}^{\rho^+ \rho_{\bar u u}^0}a_1 + M_{na;H}^{\rho^+ \rho_{\bar u u}^0} C_1]
\right. \non && \left.
- [f_{B_c} F_{fa;H}^{\rho_{\bar d d}^0\rho^+ }a_1 + M_{na;H}^{\rho_{\bar d d}^0\rho^+ } C_1] \right\}=0\;, \label{eq:rhorho}\\
{\cal M}^{H}(B_c \to \rho^+ \omega) &=& V_{cb}^* V_{ud}
\left\{[f_{B_c} F_{fa;H}^{\rho^+ \omega_{\bar u u}}a_1 + M_{na;H}^{\rho^+ \omega_{\bar u u}}]
\right. \non && \left.
+ [f_{B_c} F_{fa;H}^{ \omega_{\bar d d}\rho^+}a_1 + M_{na;H}^{\omega_{\bar d d}\rho^+ }] C_1 \right\}\;,\\
{\cal M}^{H}(B_c \to \overline{K}^{*0} K^{*+}) &=& V_{cb}^* V_{ud}
\left\{f_{B_c} F_{fa;H}^{\overline{K}^{*0} K^{*+}}a_1 +
M_{na;H}^{\overline{K}^{*0} K^{*+}} C_1 \right\}\;,\eeq
\beq
{\cal M}^{H}(B_c \to \phi K^{*+}) &=& V_{cb}^* V_{us} \left\{f_{B_c}
F_{fa;H}^{\phi K^{*+}}a_1 + M_{na;H}^{\phi K^{*+}} C_1 \right\}
\;,\\  {\cal M}^{H}(B_c \to
 K^{*+} \rho^0) &=& V_{cb}^* V_{us} \left\{f_{B_c}
F_{fa;H}^{K^{*+} \rho^0}a_1 + M_{na;H}^{K^{*+} \rho^0} C_1 \right\}\;,\\
{\cal M}^{H}(B_c \to
 K^{*0} \rho^+) &=&  \sqrt{2} {\cal M}^{H}(B_c \to
K^{*+} \rho^0)\;,\\
{\cal M}^{H}(B_c \to
 K^{*+} \omega) &=& V_{cb}^* V_{us} \left\{f_{B_c}
F_{fa;H}^{K^{*+} \omega}a_1 + M_{na;H}^{K^{*+} \omega} C_1
\right\}\;. \label{eq:ksom}  \eeq

\section{Numerical Results and Discussions}\label{sec:3}

In this section, we will calculate the branching ratios ( and
polarization fractions, relative phases) for those considered thirty
$B_c \to M_2 M_3$ decay modes. The input parameters and the wave
functions to be used are given in Appendix \ref{sec:app1}. In
numerical calculations, central values of input parameters will be
used implicitly unless otherwise stated.


\begin{table}[h]
\caption{The pQCD predictions of branching ratios($BR's$) for $B_c \to
PP$ modes. The dominant errors are induced from charm quark mass
$\rm{m_c}=1.5 \pm 0.15$ GeV, combined Gegenbauer moments $a_i$ of
related meson distribution amplitudes(See Appendix~\ref{sec:app1}
explicitly), and the chiral enhancement factors $m_0^{\pi}=1.4 \pm 0.3$ GeV
and $m_0^{K}=1.6 \pm 0.1$ GeV, respectively.} \label{tab:bcpp}
\begin{center}\vspace{-0.5cm}
\begin{tabular}[t]{l|l|l|l} \hline  \hline
Decay Modes  &                & Decay Modes &  \\
$(\Delta S=0)$ & $BR's(10^{-8}$) & $(\Delta S=1)$ & $BR's(10^{-8}$) \\
\hline
$\rm{B_c \to \pi^+ \pi^0}$ &0&
$\rm{B_c \to \pi^+ K^0}$&$4.0^{+1.0}_{-0.6}(m_c)^{+2.3}_{-1.6}(a_i)^{+0.5}_{-0.3}(m_0)$
   \\
 $\rm{B_c \to \pi^+ \eta}$ &$22.8^{+6.9}_{-4.6}(m_c)^{+7.2}_{-4.5}(a_i)^{+3.4}_{-4.2}(m_0)$&
 $\rm{B_c \to K^+ \eta}$ & $0.6^{+0.0}_{-0.0}(m_c)^{+0.6}_{-0.5}(a_i)^{+0.2}_{-0.1}(m_0)$
 \\
 $\rm{B_c \to \pi^+ \eta'}$ & $15.3^{+4.6}_{-3.1}(m_c)^{+4.8}_{-3.0}(a_i)^{+2.2}_{-2.8}(m_0)$&
$\rm{B_c \to K^+ \eta'}$ & $5.7^{+0.9}_{-0.9}(m_c)^{+1.0}_{-1.6}(a_i)^{+0.0}_{-0.3}(m_0)$
\\
$\rm{B_c \to  K^+ \overline{K}^0}$ & $24.0^{+2.4}_{-0.0}(m_c)^{+7.3}_{-6.0}(a_i)^{+6.8}_{-5.8}(m_0)$&
$\rm{B_c \to K^+ \pi^0}$& $2.0^{+0.5}_{-0.3}(m_c)^{+1.2}_{-0.8}(a_i)^{+0.3}_{-0.1}(m_0)$
\\
\hline \hline
\end{tabular}
\end{center}
\end{table}

For $B_c \to PP,PV,VP$ decays, the decay rate can be written as
\beq
\Gamma =\frac{G_{F}^{2}m^{3}_{B_c}}{32 \pi  } |{\cal A}(B_c
\to M_2 M_3)|^2\;
\eeq
where the corresponding decay amplitudes ${\cal A}$ have been
given explicitly in Eqs.~(\ref{eq:pipi0}-\ref{eq:ketap}) and
Eqs.~(\ref{eq:pirho}-\ref{eq:phik}). Using the decay amplitudes
obtained in last section, it is straightforward to calculate the
branching ratios with uncertainties as presented in
Tables~(\ref{tab:bcpp}-\ref{tab:bcvp}).

\begin{table}[h]
\caption{Same as Table~\ref{tab:bcpp} but for $B_c \to PV$ modes.}
\label{tab:bcpv}
\begin{center}\vspace{-0.8cm}
\begin{tabular}[t]{l|l|l|l} \hline  \hline
Decay Modes   &                & Decay Modes    &     \\
$(\Delta S=0)$& $BR's(10^{-7})$ & $(\Delta S=1)$ & $BR's(10^{-8})$ \\
\hline
 $\rm{B_c \to \pi^+ \rho^0}$
 &$1.7^{+0.1}_{-0.0}(m_c)^{+0.1}_{-0.2}(a_i)^{+0.6}_{-0.3}(m_0)$&
 $\rm{B_c \to K^+ \rho^0}$& $3.1^{+0.6}_{-0.8}(m_c)^{+1.2}_{-1.5}(a_i)^{+ 0.1}_{-0.2}(m_0)$  \\
 $\rm{B_c \to \overline{K}^0 K^{*+}}$
 &$1.8^{+0.7}_{-0.1}(m_c)^{+4.1}_{-2.1}(a_i)^{+0.1}_{-0.0}(m_0)$& $\rm{B_c \to K^0 \rho^+}$
 &$6.1^{+1.3}_{-1.5}(m_c)^{+2.5}_{-2.9}(a_i)^{+ 0.2}_{-0.3}(m_0)$\\
$\rm{B_c \to \pi^+ \omega}$ &
$5.8^{+1.4}_{-2.2}(m_c)^{+1.1}_{-1.3}(a_i)^{+ 0.4}_{-1.2}(m_0)$
& $\rm{B_c \to K^+ \omega}$ & $2.3^{+1.1}_{-0.3}(m_c)^{+1.8}_{-1.2}(a_i)\pm 0.1(m_0)$\\
\hline \hline
\end{tabular}
\end{center}
\end{table}

\begin{table}[t]
\caption{Same as Table~\ref{tab:bcpp} but for $B_c \to VP$ modes.}
\label{tab:bcvp}
\begin{center}\vspace{-0.8cm}
\begin{tabular}[t]{l|l|l|l} \hline  \hline
Decay Modes   &                 & Decay Modes    &  \\
$(\Delta S=0)$& $BR's(10^{-7})$ & $(\Delta S=1)$ & $BR's(10^{-8})$ \\
\hline
 $\rm{B_c \to \rho^+ \pi^0}$  &$0.5^{+0.1}_{-0.1}(m_c)^{+0.3}_{-0.2}(a_i)^{+0.2}_{-0.3}(m_0)$&
$\rm{B_c \to K^{*0}\pi^+}$ &$3.3^{+0.7}_{-0.2}(m_c)^{+0.4}_{-0.4}(a_i)^{+0.2}_{-0.1}(m_0)$\\
 $\rm{B_c \to \rho^+ \eta}$ &$5.4^{+2.1}_{-1.2}(m_c)^{+0.9}_{-1.4}(a_i)\pm 0.0(m_0)$&
 $\rm{B_c \to K^{*+}\pi^0}$& $1.6^{+0.4}_{-0.1}(m_c)^{+0.3}_{-0.1}(a_i)^{+0.1}_{-0.0}(m_0)$ \\
$\rm{B_c \to \rho^+ \eta'}$ & $3.6^{+1.4}_{-0.8}(m_c)^{+0.6}_{-0.9}(a_i)\pm 0.0(m_0)$&
$\rm{B_c \to K^{*+} \eta}$ & $0.9^{+ 0.1}_{-0.0}(m_c) ^{+0.6}_{-0.2} (a_i)\pm 0.0(m_0)$\\
$\rm{B_c \to \overline{K}^{*0} K^+}$ & $10.0^{+0.5}_{-0.6}(m_c)^{+1.7}_{-3.3}(a_i)^{+0.0}_{-0.2}(m_0)$&
$\rm{B_c \to K^{*+} \eta'}$ & $3.8\pm 1.1 (m_c)^{+1.0}_{-0.6}(a_i)\pm 0.0 (m_0)$ \\
 & &  $\rm{B_c \to \phi K^+}$  & $5.6^{+1.1}_{-0.0}(m_c)^{+1.2}_{-0.9}(a_i)^{+0.3}_{-0.0}(m_0)$ \\
\hline \hline
\end{tabular}
\end{center}
\end{table}


For $B_c \to VV$ decays, the decay rate can be written explicitly as,
\beq
\Gamma =\frac{G_{F}^{2}|\bf{P_c}|}{16 \pi m^{2}_{B_c} }
\sum_{\sigma=L,T}{\cal M}^{(\sigma)\dagger }{\cal M^{(\sigma)}}\;
\label{dr1}
\eeq
where $|\bf{P_c}|\equiv |\bf{P_{2z}}|=|\bf{P_{3z}}|$ is the momentum of either of the
outgoing vector mesons.

Based on the helicity amplitudes~(\ref{eq:amp}), we can define the
transversity amplitudes,
\beq
{\cal A}_{L}&=&-\xi m^{2}_{B_c}{\cal M}_{L}, \quad {\cal A}_{\parallel}=\xi
\sqrt{2}m^{2}_{B_c}{\cal M}_{N}, \quad {\cal A}_{\perp}=\xi
m^{2}_{B_c} \sqrt{2(r^{2}-1)} {\cal M }_{T}\;. \label{eq:ase}
\eeq
for the longitudinal, parallel, and perpendicular polarizations,
respectively, with the normalization factor
$\xi=\sqrt{G^2_{F}{\bf{P_c}} /(16\pi m^2_{B_c}\Gamma)}$ and the
ratio $r=P_{2}\cdot P_{3}/(m_{M_2}\cdot
m_{M_3})$.
These amplitudes satisfy the relation,
\beq
|{\cal A}_{L}|^2+|{\cal A}_{\parallel}|^2+|{\cal A}_{\perp}|^2=1
\eeq
following the summation in Eq.~(\ref{dr1}).

Since the transverse-helicity contributions manifest themselves in
polarization observables, we therefore define two kinds of
polarization observables, i.e., polarization fractions
$(f_{L},f_{||},f_{\perp})$ and relative phases $(\phi_{||},\phi_{\perp})$ as~\cite{phikst},
\beq
f_{L(||,\perp)}= \frac{|{\cal A}_{L(||,\perp)}|^2}{|{\cal
A}_L|^2+|{\cal A}_{||}|^2+|{\cal A}_{\perp}|^2},\;\;\;\;\;\;\;\;
\phi_{||(\perp)} \equiv \arg \frac{{\cal A}_{||(\perp)}}{{\cal A}_{L}}\;;
\label{eq:pf}
\eeq
It should be noted that the final results of
relative phases will plus one value, i.e., $\pi$, due to  an
additional minus sign in the definition of ${\cal A}_L$.

We also define another two quantities reflecting the effects of
CP-violating asymmetries indirectly~\cite{phikst,bry},
\beq
\Delta \phi_{||}&=&
\frac{\bar{\phi}_{||}-\phi_{||}}{2}\;, \qquad
\Delta\phi_{\perp}=\frac{\bar{\phi}_\perp-\phi_\perp-\pi}{2}\;,
\eeq
where $\bar{\phi}_{||}$ and $\bar{\phi}_\perp$ are the
CP-conjugated relative phases corresponding to $\phi_{||}$ and
$\phi_{\perp}$, respectively.

With the complete decay amplitudes, by employing Eq.(\ref{dr1})
and the input parameters and wave functions as given in Appendix
\ref{sec:app1}, we will present the pQCD predictions for
CP-averaged branching ratios, longitudinal polarization fractions
and relative phases of the considered decays with errors as shown
in Tables~\ref{tab:bcvv} and~\ref{tab:bcrp}.

\begin{table}[htb]
\caption{The pQCD predictions of branching ratios($BR's$) and
longitudinal polarization fractions($LPF's$) for $B_c \to V V$ modes.
} \label{tab:bcvv}
\begin{center}\vspace{-0.4cm}
\begin{tabular}[t]{l|c|c} \hline  \hline
Decay Modes  & $BR's(10^{-7})$ & $LPF's$ (\%)   \\
\hline
  $\rm{B_c \to \rho^+ \rho^0}$
 &0 & $-$  \\
 $\rm{B_c \to \rho^+ \omega}$
 &$10.6^{+3.2}_{-0.2}(m_c)^{+2.1}_{-0.2}(a_i)$& $92.9^{+1.6}_{-0.1}(m_c)^{+1.2}_{-0.1}(a_i)$ \\
 $\rm{B_c \to \overline{K}^{*0} K^{*+}}$
 &$10.0^{+0.6}_{-0.4}(m_c)^{+8.1}_{-4.8}(a_i)$& $92.0^{+0.5}_{-0.4}(m_c)^{+3.6}_{-7.1}(a_i)$ \\
 \hline
 $\rm{B_c \to K^{*0} \rho^+}$
 & $0.6^{+0.0}_{-0.0}(m_c)^{+0.2}_{-0.1}(a_i)$& $94.9^{+0.2}_{-0.2}(m_c)^{+2.0}_{-1.4}(a_i)$  \\
 $\rm{B_c \to K^{*+} \rho^0}$
 & $0.3^{+0.0}_{-0.0}(m_c)^{+0.1}_{-0.1}(a_i)$& $94.9^{+0.2}_{-0.2}(m_c)^{+1.3}_{-1.4}(a_i)$ \\
 $ \rm{B_c \to K^{*+} \omega}$
 & $0.3^{+0.0}_{-0.0}(m_c)^{+0.0}_{-0.2}(a_i)$& $94.8^{+0.3}_{-0.2}(m_c)^{+1.1}_{-1.2}(a_i)$ \\
 $\rm{B_c \to \phi K^{*+}}$
 & $0.5^{+0.0}_{-0.1}(m_c)^{+0.1}_{-0.3}(a_i)$& $86.4^{+0.0}_{-1.4}(m_c)^{+4.9}_{-9.0}(a_i)$ \\ \hline \hline
\end{tabular}
\end{center}
\end{table}

\begin{table}[htb]
\caption{The pQCD predictions of relative phases for $B_c \to V V$
modes.} \label{tab:bcrp}
\begin{center}\vspace{-0.5cm}
\begin{tabular}[t]{l|c|c|c|c} \hline  \hline
Decay Modes  & $\phi_{||}$ (rad) & $\phi_{\perp}$ (rad)& $\Delta \phi_{||}$ & $\Delta\phi_{\perp}$ \\
\hline
  $\rm{B_c \to \rho^+ \rho^0}$
 &$-$& $-$ & $-$ & $-$  \\
 $\rm{B_c \to \rho^+ \omega}$
 &$3.86^{+0.31}_{-0.26}(m_c)^{+0.25}_{-0.19}(a_i)$& $4.43^{+0.16}_{-0.17}(m_c)^{+0.25}_{-0.19}(a_i)$& $0$ & $-\pi/2$ \\
 $\rm{B_c \to \overline{K}^{*0} K^{*+}}$
 &$3.68^{+0.18}_{-0.13}(m_c)^{+0.48}_{-0.21}(a_i)$& $3.76^{+0.16}_{-0.00}(m_c)^{+0.48}_{-0.20}(a_i)$& $0 $& $-\pi/2$ \\
\hline
 $\rm{B_c \to K^{*0} \rho^+}$
 & $4.11^{+0.17}_{-0.20}(m_c)^{+0.30}_{-0.20}(a_i)$& $4.20^{+0.14}_{-0.05}(m_c)^{+0.30}_{-0.21}(a_i)$& $0$ & $-\pi/2$ \\
 $\rm{B_c \to K^{*+} \rho^0}$
 & $4.11^{+0.17}_{-0.20}(m_c)^{+0.30}_{-0.20}(a_i)$& $4.20^{+0.14}_{-0.05}(m_c)^{+0.30}_{-0.21}(a_i)$& $0$ &  $-\pi/2$\\
 $ \rm{B_c \to K^{*+} \omega}$
 & $4.15^{+0.13}_{-0.25}(m_c)^{+0.25}_{-0.25}(a_i)$& $4.23^{+0.11}_{-0.09}(m_c)^{+0.26}_{-0.24}(a_i)$& $0$ & $-\pi/2$ \\
 $\rm{B_c \to \phi K^{*+}}$
 & $3.80^{+0.25}_{-0.34}(m_c)^{+0.44}_{-0.20}(a_i)$& $3.89^{+0.22}_{-0.19}(m_c)^{+0.43}_{-0.21}(a_i)$& $0$ & $-\pi/2$\\ \hline \hline
\end{tabular}
\end{center}
\end{table}

\begin{table}[htb]
\caption{The pQCD predictions of branching ratios for
$B_c \to \phi K^+$ and $B_c \to \overline{K}^{(*)0} K^{(*)+}$ modes.
As a comparison, the numerical results as given in Ref.~\cite{ekou09:ncbc}
are also listed in the last two columns. }
\label{tab:bcbr}
\begin{center}\vspace{-0.6cm}
\begin{tabular}[t]{l|l|l|l} \hline  \hline
Channels  &  pQCD Predictions  &  SU(3) Symmetry & OGE model \\
\hline
 $ \rm{Br}(B_c \to \phi K^{+}) $
 & $5.6^{+1.1}_{-0.0}(m_c)^{+1.2}_{-0.9}(a_i) \times 10^{-8}$& $ {\cal O} (10^{-7} \sim 10^{-8})$& $5 \times 10^{-9}$ \\
  $\rm{Br}(B_c \to \overline{K}^{0} K^{+})$
 &$2.4^{+0.2}_{-0.0}(m_c)^{+0.7}_{-0.6}(a_i) \times 10^{-7}$ & ${\cal O} (10^{-6})$ & $6.3 \times 10^{-8}$ \\
$\rm{Br}(B_c \to \overline{K}^{0} K^{*+})$
&$1.8^{+0.7}_{-0.1}(m_c)^{+4.1}_{-2.1}(a_i)\times 10^{-7}$& $-$& $-$ \\
 $\rm{Br}(B_c \to \overline{K}^{*0} K^{+})$
 &$1.0\pm 0.1(m_c)^{+0.2}_{-0.3}(a_i) \times 10^{-6}$& ${\cal O} (10^{-6})$& $9.0 \times 10^{-8}$ \\
 $\rm{Br}(B_c \to \overline{K}^{*0} K^{*+})$
 & $1.0^{+0.1}_{-0.0}(m_c)^{+0.8}_{-0.5}(a_i) \times 10^{-6}$& ${\cal O} (10^{-6})$ &$9.1 \times 10^{-8}$ \\
 \hline \hline
\end{tabular}
\end{center}
\end{table}

Based on the pQCD predictions as given in Tables I -V, we have the following remarks:
\begin{itemize}
\item
Among considered pure annihilation $B_c\to PV/VP, VV $ decays, the
pQCD predictions for the CP-averaged branching ratios for those
$\Delta S = 0$ processes are much larger than those of $\Delta S =1$
channels ( one of the two final state mesons is the $K^{(*)}$ meson
), which are mainly due to the large CKM factor $|V_{ud}/V_{us}|^2
\sim 19$. For $B_c \to \pi^+ \pi^0$, $\rho^+ \rho^0$ decays, the
contributions from $\bar u u$ and $\bar d d$ components cancel each
other exactly and result in the zero branching ratios. In fact,
these two channels are forbidden, even if with final state
interactions. Simply, two pions can not form an s wave isospin 1
state, because of Bose-Einstein statics.
 Any other nonzero data for these two channels may
indicate the effects of exotic new physics.

\item
There is no CP violation for all these decays within the standard model, since there
is only one kind of tree operators involved in the decay amplitude of
all considered $B_c$ decays, which can be seen from Eq.~(\ref{eq:amt}).

\item
The pQCD predictions for the branching ratios of considered $B_c$ decays vary in the
range of $10^{-6}$ ( for $B_c \to \overline{K}^{*0} K^{+},\overline{K}^{*0} K^{*+}$ and $\rho^+ \omega$ decays)
to  $10^{-8}$ ( for most $\Delta S=1$ $B_c$ decays).
The $B_c$ decays with the branching ratio of $10^{-6}$ can be measured at
the LHC experiment \cite{ekou09:ncbc}.

\item
As mentioned in the introduction, the authors of Ref.~\cite{ekou09:ncbc} studied
many pure annihilation $B_c$ decays by employing the SU(3) flavor symmetry and
the OGE model respectively, and presented their numerical estimates
for the branching ratios of $B_c \to \phi K^+, \overline{K}^0 K^+,
\overline{K}^{*0} K^+$ and $\overline{K}^{*0} K^{*+}$ decays. As a comparison, we
show in Table VI the pQCD predictions and the results as given in
Ref.~\cite{ekou09:ncbc} for relevant channels.
From Table VI, one can see easily that the pQCD predictions basically agree with the
results obtained based  on the $SU(3)$ flavor symmetry.

\item
For $B_c \to (\pi^+, \rho^+) (\eta, \eta')$ decays, the relevant final state
mesons  contain the same component $\bar u u + \bar d d$, they therefore have the similar branching
ratios. The small differences among their branching ratios mainly come from the different
mixing coefficients, i.e., $\cos\phi$ and $\sin\phi$.

\item
For $B_c \to K^+ \eta^{(')}$ decays, however, one finds that
$Br(B_c \to K^+ \eta^\prime) \sim 10 \times Br(B_c \to K^+ \eta)$,
which is rather different from  the pattern of
$Br(B_c \to \pi^+ \eta ) \sim Br(B_c \to \pi^+ \eta^\prime)$
and $Br(B_c \to \rho^+ \eta) \sim Br(B_c \to \rho^+ \eta^\prime)$.
This large difference can be understood as follows: For the $\Delta S= 1$ processes,
both $\eta_q$ and $\eta_s$ will contribute to $B_c \to
K^+ \eta$ and  $K^+ \eta^\prime$ decays but with an opposite sign for $\eta_q$ and
$\eta_s$ term, as well as different coefficients.
Which results in a destructive interference between $\eta_q$ and $\eta_s$ component
for $B_c \to K^+ \eta$, but a constructive interference
for $B_c \to K^+ \eta^\prime$. This situation
is very similar with that for the $B \to K \eta$ and $K \eta^\prime$ decays
~\cite{Amsler08:pdg,Barberio08:hfag,xiao08:keta}.

\item
Unlike $B_c \to K^+ \eta^{(')}$ decays, $Br(B_c \to K^{*+} \eta^\prime)\approx 4
Br(B_c \to K^{*+} \eta) \sim 3.8 \times 10^{-8}$. The reason is that
both of them are mainly determined by the factorizable contributions of $\eta_s$ term.

\item
For $B_c \to VV$ decays, we can find that (a) the branching ratios
are in order of ${\cal O}(10^{-8} \sim 10^{-7})$ except for
$\rm{Br(B_c \to \overline{K}^{*0} K^{*+})}$ and  $\rm{Br(B_c \to \rho^+ \omega)} \sim 10^{-6}$;
(b) the longitudinal polarization fractions are around $95\%$
within the theoretical errors except for $B_c \to \phi K^{*+}$ ( $\sim 86\%$)
and play the dominant role.

\item
According to the discussions in Ref.~\cite{ekou09:ncbc},
there are some simple relations among some decay channels in the limit of
exact $SU(3)$ flavor symmetry.
For $B_c \to PP$ decays, such relations are
\beq
A(B_c\to K^0\pi^+)=\sqrt{2}A(B_c\to K^+\pi^0)= \lambda A(B_c\to
K^+\bar{K}^0)\;,
\label{kpi}
\eeq
where  $\lambda=V_{us}/V_{ud}\approx 0.2$.
For $B_c \to VP/PV$ and $B_c \to VV$ decays, the relations read
\beq
A(B_c\to K^{*0}\pi^+) &=& \sqrt{2}A(B_c\to K^{*+}\pi^0)
=\lambda A(B_c\to \bar{K}^{*0} K^+), \\
A(B_c\to \rho^+ K^0) &=& \sqrt{2}A(B_c\to \rho^0 K^+) = \lambda
A(B_c\to K^{*+} \bar{K}^0), \\
(-1)^\ell A(B_c^+\to \rho^+ K^{*0})
 &=& (-1)^\ell \sqrt{2} A(B_c^+\to \rho^0 K^{*+}) = \lambda A(B_c\to K^{*+} \bar{K}^0)
\label{kpip}\eeq
where $\ell=0,1,2$
\footnote{Here, since the
longitudinal contributions dominate the $B_c \to K^{*0} \rho^+$ decay,
we use its longitudinal part (i.e., $\ell=0$ ) to compare with the decay amplitude
of $B_c\to K^{*+} \bar{K}^0$ decay.}.
From our pQCD calculations, we
notice  that the first equality of each of the above relations
(\ref{kpi}-\ref{kpip}) are valid in isospin symmetry. They hold
exactly in our numerical calculations. The second equality of each
relations are only valid at exact SU(3) symmetry thus they are
violated   at  the order of SU(3) breaking effect in our calculations.

\item
Since the LHC experiment can measure the $B_c$ decays with a
branching ratio at $10^{-6}$ level, our pQCD predictions for
the branching ratios of $B_c \to \overline{K}^{*0} K^{+}$, $\overline{K}^{*0} K^{*+}$ and $\rho^+ \omega$
decays could be tested in the forthcoming LHC experiments.

\item
For most considered pure annihilation $B_c$ decays, it is hard to observe them
even in LHC due to their tiny decay rate.
Their observation at LHC, however, would mean  a large non-perturbative
contribution or a signal for new physics beyond the SM.

\item
It is worth of stressing that the theoretical predictions in
the pQCD approach still have large theoretical errors induced
by the still large uncertainties of many input parameters.
Any progress in reducing the error of input parameters, such as
the Gegenbauer moments $a_i$ and the charm quark mass $m_c$, will help us
to improve the precision of the pQCD predictions.

\end{itemize}


\section{Summary}\label{sec:sum}

In short, we studied the two-body charmless hadronic $B_c \to PP, PV/VP, VV$
decays by employing the pQCD factorization approach based on the
$k_T$ factorization theorem. These considered decay channels can
occur only via the annihilation diagram and they will provide an important
testing ground for the magnitude of the annihilation contribution.

The pQCD predictions for CP-averaged branching ratios,
longitudinal polarization fractions and relative phases are
displayed in Tables~(\ref{tab:bcpp}-\ref{tab:bcrp}).
From our numerical evaluations and phenomenological analysis, we
found the following results:
\begin{itemize}
\item
The pQCD predictions for the branching ratios
vary in the range of $10^{-6}$ to $10^{-8}$, basically agree with
the predictions obtained by using the exact SU(3) flavor symmetry.
The $B_c\to \overline{K}^{*0} K^{+} $ and other decays with a decay rate
at $10^{-6}$ or larger  could be measured at the LHC experiment.

\item
For $B_c \to PV/VP, VV$ decays, the branching ratios of  $\Delta S= 0$ processes
are basically larger than those of $\Delta S =1$ ones.
Such differences are mainly induced  by
the CKM factors involved: $V_{ud}\sim 1 $ for the former decays
while $V_{us}\sim 0.22$ for the latter ones.

\item
Analogous to $B \to K \eta^{(\prime)}$ decays, we find
$Br(B_c \to K^+ \eta^\prime) \sim 10 \times Br(B_c \to K^+ \eta)$.
This large difference can be understood
by the destructive and constructive interference
between the $\eta_q$ and $\eta_s$ contribution to
the $B_c \to K^+ \eta$ and $B_c \to K^+ \eta^\prime$ decay.

\item
For $B_c \to VV$ decays, the longitudinal polarization fractions are around $95\%$
except for $B_c \to \phi K^{*+}$ ( $f_L \sim 86\%$) and play the dominant role.

\item
Because only tree operators are involved, the CP-violating asymmetries for
these considered $B_c$ decays are absent naturally.

\item
The pQCD predictions still have large theoretical uncertainties,
induced by the uncertainties of input parameters.

\item
We here calculated the branching ratios and other physical
observables of the pure annihilation $B_c$ decays by employing the pQCD approach.
We do not consider the possible long-distance (LD) contributions, such as the
re-scattering effects, although they may be large and affect the theoretical
predictions. It is beyond the  scope of this work.

\end{itemize}

\begin{acknowledgments}

X.~Liu is very grateful to Dr. Jun-Feng~Sun and Dr. Xian-Qiao~Yu for
helpful discussions. This work is supported by the
National Natural Science Foundation of China under Grant No.10975074, 10625525
and 10735080, and supported by Project on Graduate Students'
Education and Innovation of Jiangsu Province, under Grant No.
${\rm CX09B_{-}297Z}$.

\end{acknowledgments}


\begin{appendix}

\section{Input parameters and distribution amplitudes} \label{sec:app1}

The masses~({\rm GeV}), decay constants~({\rm GeV}), QCD
scale~({\rm GeV}) and $B$ meson lifetime are
\beq
 \Lambda_{\overline{\mathrm{MS}}}^{(f=4)} &=& 0.250, \quad m_W = 80.41, \quad  m_{B_c} = 6.286, \quad  f_{B_c} = 0.489,  \non
 m_\phi &=& 1.02,\;\;\;\; \quad f_{\phi} = 0.231,\;\;\;\quad
 f_{\phi}^T = 0.200, \; m_{K^*}=0.892,
   \non
f_{K^*} &=& 0.217, \quad f_{K^*}^T = 0.185,\;\; \quad m_{\rho} =
0.770, \;\;\quad f_{\rho}= 0.209,
 \non
   f^T_{\rho}&=& 0.165, \;\quad m_{\omega}=0.782, \;\;\;\quad f_{\omega}= 0.195,
  \; \quad f_{\omega}^T =0.145, \non
     m^\pi_{0}&=& 1.4,\;\;\;\; \quad m_0^K=1.6,\;\;\;\; \quad m_0^{\eta_q}= 1.08,
  \;\; \quad m_0^{\eta_s} =1.92, \non
    m_b &=& 4.8, \;\;\;\;\;\;\quad f_{\pi}= 0.131, \;\;\quad f_K = 0.16, \;\;\quad
    \tau_{B_c^+}= 0.46\; ps\;.
 \label{para}
\eeq

For the CKM matrix elements, here we adopt the Wolfenstein
parametrization for the CKM matrix, and take $A=0.814$ and
 $\lambda=0.2257$, $\bar{\rho}=0.135$ and $\bar{\eta}=0.349$ \cite{Amsler08:pdg}.

The twist-2 pseudoscalar meson distribution amplitude
$\phi_P^A$($P=\pi,K$) , and the twist-3 ones $\phi_P^P$ and
$\phi_P^T$ have been parametrized
as~\cite{PseudoscalarWV,Ball99:Pseudoscalar,PseudoscalarUpdate},
\beq \phi_{P}^A(x) &=& \frac{f_{P}}{2\sqrt{2N_c}}\, 6x(1-x) \left[1
+ a_1^{P} C_1^{3/2}(2x-1) +
a_2^{P}C_2^{3/2}(2x-1)+a_4^{P}C_4^{3/2}(2x-1)\right] \;, \\
\phi^P_{P}(x) &=& \frac{f_{P}}{2\sqrt{2N_c}}\, \bigg[ 1
+\left(30\eta_3 -\frac{5}{2}\rho_{P}^2\right) C_2^{1/2}(2x-1) \non &
& \hspace{35mm} -\, 3\left\{ \eta_3\omega_3 +
\frac{9}{20}\rho_{P}^2(1+6a_2^{P}) \right\} C_4^{1/2}(2x-1)
\bigg]\;,\\ \phi^T_{P}(x) &=& \frac{f_{P}}{2\sqrt{2N_c}}\,
(1-2x)\bigg[ 1 + 6\left(5\eta_3 -\frac{1}{2}\eta_3\omega_3 -
\frac{7}{20}
      \rho_{P}^2 - \frac{3}{5}\rho_{P}^2 a_2^{P} \right)
(1-10x+10x^2) \bigg]\;,\ \ \ \ \eeq with the Gegenbauer moments
$a_1^{\pi}=0, a_1^{K}=0.17 \pm 0.17, a_2^{P}= 0.115 \pm 0.115,
a_4^{P}=-0.015$, the mass ratio
$\rho_{\pi(K)}=m_{\pi(K)}/m_{0}^{\pi(K)}$ and $\rho_{\eta_{q(s)}}=2
m_{q(s)}/m_{qq(ss)}$, and the Gegenbauer polynomials $C_n^{\nu}(t)$,
\begin{eqnarray}
 C_2^{1/2}(t)\,& =&\, \frac{1}{2} \left(3\, t^2-1\right) \;,
C_4^{1/2}(t)\, =\, \frac{1}{8} \left(3-30\, t^2+35\, t^4\right) \;,
\nonumber\\
C_1^{3/2}(t)\, &=&\, 3\, t \;, \; \; \;  C_2^{3/2}(t)\, =\,
\frac{3}{2} \left(5\, t^2-1\right) \;, \;\; C_4^{3/2}(t) \,=\,
\frac{15}{8} \left(1-14\, t^2+21\, t^4\right) \;.
\end{eqnarray}
In the above distribution amplitudes for kaon, the momentum fraction
$x$ is carried by the $s$ quark. For both the pion and kaon, we
choose $\eta_3=0.015$ and $\omega_3=-3$
\cite{PseudoscalarWV,Ball99:Pseudoscalar}.

The twist-2 distribution amplitudes for the longitudinally and
tranversely polarized vector meson can be parameterized as: \beq
\phi_{V}(x)&=&\frac{3f_{V}}{\sqrt{6}} x
(1-x)\left[1+a_{1V}^{||}C^{3/2}_1(2x-1)+ a_{2V}^{||}C_2^{3/2}
(2x-1)\right]\;,\label{eq:lda}\\
\phi_{V}^T(x)&=&\frac{3f^T_{V}}{\sqrt{6}} x
(1-x)\left[1+a_{1V}^{\perp}C^{3/2}_1(2x-1)+ a_{2V}^{\perp}C_2^{3/2}
(2x-1)\right]\;,\label{eq:tda} \eeq Here $f_{V}$ and $f_V^T$ are the
decay constants of the vector meson with longitudinal and tranverse
 polarization, respectively. The Gegenbauer moments have been studied extensively in the
literatures \cite{rho,previousvectorwf}, here we adopt the following
values from the recent updates~\cite{adoptedvectorwf,LCSRBZ,twist3}:
\beq a_{1K^*}^{||}&=&0.03\pm0.02,
a_{2\rho}^{||}=a_{2\omega}^{||}=0.15\pm0.07,
a_{2K^*}^{||}=0.11\pm0.09,a_{2\phi}^{||}=0.18\pm0.08\\
a_{1K^*}^\perp&=&0.04\pm0.03,
a_{2\rho}^{\perp}=a_{2\omega}^{\perp}=0.14\pm0.06,
a_{2K^*}^{\perp}=0.10\pm0.08,a_{2\phi}^{\perp}=0.14\pm0.07 \eeq

The asymptotic forms of the twist-3 distribution amplitudes
$\phi^{t,s}_V$ and $\phi_V^{v,a}$ are \cite{Li05:kphi}: \beq
\phi^t_V(x) &=& \frac{3f^T_V}{2\sqrt 6}(2x-1)^2,\;\;\;\;\;\;\;\;\;\;\;
  \hspace*{0.5cm} \phi^s_V(x)=-\frac{3f_V^T}{2\sqrt 6} (2x-1)~,\\
\phi_V^v(x)&=&\frac{3f_V}{8\sqrt6}(1+(2x-1)^2),\;\;\; \ \ \
 \phi_V^a(x)=-\frac{3f_V}{4\sqrt6}(2x-1).
\eeq

\section{Related hard functions} \label{sec:app2}
In this section, we group the functions which appear in the
factorization formulae.

The functions $h$  in the decay amplitudes consist of two parts: one
is the jet function $S_t(x_i)$ derived by the threshold
re-summation\cite{Li02:resum}, the other is the propagator of virtual quark
and gluon. They are defined by
\begin{eqnarray}
h_{fa}(x_3,x_2,b_3,b_2)&=&(\frac{i\pi}{2})^2
S_t(x_2)\Big[\theta(b_3-b_2)H_0^{(1)}(\sqrt{x_2}M_{B_c}b_3)J_0(\sqrt
{x_2}M_{B_c}b_2)\nonumber\\
&&+\theta(b_2-b_3)H_0^{(1)}(\sqrt {x_2}M_{B_c}b_2)J_0(\sqrt
{x_2}M_{B_c}b_3)\Big]H_0^{(1)}(\sqrt{x_2 x_3}M_{B_c}b_3),
\\
h_{na}^{c(d)}(x_2,x_3,b_1,b_2)&=&\frac{i\pi}{2}\left[\theta(b_1-b_2)H^{(1)}_0(\sqrt
{x_2(1-x_3)}M_{B_c}b_1)J_0(\sqrt {x_2(1-x_3)}M_{B_c}b_2)\right. \nonumber\\
&&\;\;\left.
+\theta(b_2-b_1)H^{(1)}_0(\sqrt{x_2(1-x_3)}M_{B_c}b_2)J_0(\sqrt
{x_2(1-x_3)}M_{B_c}b_1)\right]\nonumber\\
&&\;\;\;\times
\left\{\begin{array}{ll}\frac{i\pi}{2}H^{(1)}_0(\sqrt{|F_{c(d)}^2|}M_{B_c}b_1),&
F_{c(d)}<0\\
K_0(\sqrt {F_{c(d)}}M_{B_c}b_1),& F_{c(d)}>0\end{array}\right. ,
\end{eqnarray}
where \beq F_{c}&=& (r_c-x_2)(1-x_3)+r_c^2\;,\;\;\;\;\;\;\;\;\; F_d=
r_b^2-(1-r_c-x_2)x_3 \;, \eeq and $H_0^{(1)}(z) = \mathrm{J}_0(z) +
i\, \mathrm{Y}_0(z)$.

The hard scales are chosen as \begin{eqnarray}
t_a&=&\mbox{max}\{\sqrt {x_2}M_{B_c},1/b_2,1/b_3\},\\
t_b&=&\mbox{max}\{\sqrt{1-x_3}M_{B_c},1/b_2,1/b_3\},\\
t_c&=&\mbox{max}\{\sqrt {x_2(1-x_3)}M_{B_c},
\sqrt{|(r_c-x_2)(1-x_3)+r_c^2|}M_{B_c},1/b_1,1/b_2\},\\
t_d&=&\mbox{max}\{\sqrt{x_2(1-x_3)}M_{B_c},\sqrt{|r_b^2-(1-r_c-x_2)x_3|}M_{B_c},1/b_1,1/b_2\}.
\end{eqnarray} They are given as the maximum energy scale appearing in each diagram to kill the large
logarithmic radiative corrections.

The $S_t$ re-sums the threshold logarithms $\ln^2x$ appearing in the
hard kernels to all orders and it has been parameterized as
  \begin{eqnarray}
S_t(x)=\frac{2^{1+2c}\Gamma(3/2+c)}{\sqrt \pi
\Gamma(1+c)}[x(1-x)]^c,
\end{eqnarray}
with $c=0.4 \pm 0.1$. In the nonfactorizable contributions, $S_t(x)$
gives a very small numerical effect to the amplitude~\cite{L4}.
Therefore, we drop $S_t(x)$ in $h_{na}$.

The evolution factors $E_{fa}$ and $E_{na}$ entering in the
expressions for the matrix elements (see section III) are given by
\begin{eqnarray}
E_{fa}(t)&=&\alpha_s(t)
 \exp[-S_2(t)-S_3(t)],\\
E_{na}(t)&=&\alpha_s(t) \exp[-S_B(t)-S_2(t)-S_3(t)]|_{b_2=b_3},
\end{eqnarray}
in which the Sudakov exponents are defined as
\begin{eqnarray}
S_B(t)&=&s\left(r_c \frac{M_{B_c}}{\sqrt
2},b_1\right)+\frac{5}{3}\int^t_{1/b_1}\frac{d\bar \mu}{\bar
\mu}\gamma_q(\alpha_s(\bar \mu)),\\
S_2(t)&=&s\left(x_2\frac{M_{B_c}}{\sqrt
2},b_2\right)+s\left((1-x_2)\frac{M_{B_c}}{\sqrt
2},b_2\right)+2\int^t_{1/b_2}\frac{d\bar \mu}{\bar
\mu}\gamma_q(\alpha_s(\bar \mu)),
\end{eqnarray}
 with the quark
anomalous dimension $\gamma_q=-\alpha_s/\pi$. Replacing the
kinematic variables of $M_2$ to $M_3$ in $S_2$, we can get the
expression for $S_3$. The explicit forms for the function $s(Q,b)$
are defined in the Appendix A in Ref.~\cite{Lu01:pipi}.%

\end{appendix}


\end{document}